\documentclass[review]{elsarticle}
\makeatletter
\def\ps@pprintTitle{
	\let\@oddhead\@empty
	\let\@evenhead\@empty
	\def\@oddfoot{\reset@font\hfil\thepage\hfil}
	\let\@evenfoot\@oddfoot
}
\makeatother
 \biboptions{comma,sort&compress}

\usepackage{graphicx}
\usepackage{amsmath}
\usepackage{here}
\usepackage{dsfont}
\usepackage{multirow}
\usepackage{cuted}
\usepackage{hhline}

\def\mqq{\left\langle m_q \overline{q}q\right\rangle}

\def\trho{\boldsymbol{\varrho}}

\usepackage[dvips]{epsfig}

\begin{document}
\markboth{Stephan Narison, Montpellier (FR)}{ }
\begin{frontmatter}

\title{Strengthening the Bridge Between Chiral Lagrangians and QCD Sum-Rules\,\tnoteref{invit}}
\tnotetext[invit]{Talk given at QCD25 - 40th anniversary of the QCD-Montpellier Conference. }
\author{J.~Ho%\corref{cor1}
}
\address{Department of Physics, Dordt University, Sioux Center, Iowa, 51250, USA}
\ead{jason.ho@dordt.edu}
\author{Amir~H.~Fariborz}
\address{Department of Mathematics/Physics,  SUNY Polytechnic Institute, Utica, NY 13502, U.S.A.
}
\ead{fariboa@sunypoly.edu}
\author{T.G.~Steele}
\address{Department of Physics and
	Engineering Physics, University of Saskatchewan, Saskatoon, SK,
	S7N 5E2, Canada}
\ead{Tom.Steele@usask.ca}

%%
%%%%%%%%%%%%%%%%%%%%%%%%%%%%%%%%%%%%%%%%%%%%%%%%%

\date{\today}
\begin{abstract}
%\noindent
Previous work has shown that mesonic fields in chiral Lagrangians can be systematically connected to quark-level operators in QCD sum rules through chiral-symmetry constrained and energy-independent scale factor matrices. This framework yields universal scale factors associated with each chiral nonet, whether composed of quark-antiquark ($q\bar{q}$) or four-quark ($qq\bar{q}\bar{q}$) operators. Building on the demonstrated scale-factor universality for the $K_0^*$ isodoublet and $a_0$ isotriplet scalar mesons, we develop a revised Gaussian QCD sum-rule methodology that extends the analysis to higher-dimensional isospin sectors. To access nonperturbative information about resonances arising from final-state interactions, we introduce a background-resonance interference approximation. This approximation successfully reproduces both $\pi K$ scattering amplitude data and $\pi\eta$ scattering predictions. It also motivates new resonance models that enhance the scale-factor analysis linking chiral Lagrangians to QCD sum rules. Within this refined framework, we explore the scale factors for the $K_0^*$ and $a_0$ mesons across a sequence of increasingly detailed resonance models.

\end{abstract}
\end{frontmatter}
%%%%%%%%%%%%%%%%%%%%%%%%%%%%%%%%%%
\newpage
%%%%%%%%%%%%%%%%%%%%%%%%%%%%%%%%%%
\section{Introduction}
%%%%%%%%%%%%%%%%%%%%%%%%%%%%%%%%%%%
The fundamentally nonperturbative nature of the strong interaction makes it difficult to explore the low-energy regime of QCD \cite{PDG,PKZ_review}. The unique properties of the light scalar meson states is one long-standing example of this. Experimentally, this region contains complex kinematic effects and resonance shapes, and the light hadronic states of interest can have very small lifetimes. Theoretically, the large nonpertubative effects can be difficult to understand and predict. Two specific theoretical approaches that have long been used to analyze this challenging energy regime are chiral Lagrangians \cite{GLSM_pieta,GLSM_piK,Parganlija:2012fy,GLSM_pipi,GLSM,GLSM_inst,Giacosa:2006tf,00_BFMNS,00_BFS61,NLCL_kappa,Carter:1995zi,Ko:1994en}, which use effective hadronic-level operators and model parameters fixed by experimental data, and QCD sum-rules \cite{SVZ,Reinders:1984sr,Narison:2002woh}, which model hadronic bound states using constituent field operators. Each of these methods have their own strengths and limitations, but may be complementary in their approach. One limitation of the QCD sum-rule methodology is in describing resonant states with complex mixing or exotic substructure; in our previous work \cite{CLQCDSR_2016,CLQCDSR:2019_Proc,CLQCDSR_2020} we showed that this can be overcome by connecting the predictions from QCD sum-rules to the experimentally-driven modeling of chiral Lagrangians. A second limitation to the QCD sum-rule methodology is one we wish to explore in this review of our recent work \cite{ExtBridge}. A traditional analysis methodology for QCD sum-rules involves modeling the unknown hadronic dynamics with a parameterized narrow resonance with a continuum contribution \cite{SVZ}. However, for the broad and possibly overlapping scalar meson states, a different model that more precisely replicates the nature of the resonance may be required. The distinct approaches of the chiral Lagrangian and QCD sum-rule methodologies are complementary; chiral Lagrangians are able to model the resonance structure of scalar meson states, while QCD sum-rules provide a direct connection to the constituent quark structure of the mesons. Here, we use chiral Lagrangians (specifically a generalized linear sigma model) to build sophisticated resonances models which we implement in a Gaussian sum-rule analysis \cite{ExtBridge}, building on previous work connecting the two methodologies \cite{CLQCDSR_2016,CLQCDSR:2019_Proc,CLQCDSR_2020}.

\section{Methodology}
\subsection{A review of the scale factor methodology}
We briefly review the methodology developed in our previous works \cite{CLQCDSR_2016,CLQCDSR:2019_Proc,CLQCDSR_2020}. Our framework emerges from the chiral symmetry inherent in linear and non-linear chiral Lagrangian models. Two chiral nonets are defined, with $M$ representing two-quark ($\bar{q}{q}$) structures, and $M'$ representing four-quark ($\bar{q}{q}\bar{q}{q}$) structures. These chiral nonets have identical chiral transformation properties,
\begin{gather}
    M \rightarrow U_L\, M \,U_R^\dagger, \quad
    M' \rightarrow U_L\, M' \,U_R^\dagger,
\end{gather}
but due to their underlying structure, distinct $U_A(1)$ transformations:
\begin{gather}
    M \rightarrow  e^{2i\nu}M ,\quad
    M' \rightarrow e^{-4i\nu} M'.
    \label{eq:chiral-symmetry-trans}
\end{gather}
These chiral nonets are built from bare, unmixed fields representing the mesonic scalar $\{S,S'\}$ and pseudoscalar $\{\phi,\phi'\}$ fields in the chiral Lagrangian formalism,
\begin{gather}
    M = S+ i\phi,\quad M' = S' + i\phi',
\end{gather}
where $S$ and $S'$ written in terms of the chiral Langrangian bare fields are 
\begin{equation}
    \!S=
        \begin{pmatrix}
            S_1^1 & a_0^+ & \kappa^+  \\
            a_0^- & S_2^2 & \kappa^0 \\
            \kappa^- & {\bar \kappa}^0 & S_3^3
        \end{pmatrix}, 
        ~
        S' =
        \begin{pmatrix} 
            {S'}_1^1 & {a'}_0^+ & {\kappa'}^+  \\
            {a'}_0^- & {S'}_2^2 & {\kappa'}^0 \\
            {\kappa'}^- & {\bar {\kappa'}}^0 & {S'}_3^3 \\
        \end{pmatrix}.
        \label{eq:CL-scalarnonets}
\end{equation}
These chiral nonets may also be expressed in terms of quark-level operators. The quark-antiquark operators can be used to represent the scalar nonet, for example,  
\begin{equation}
    \left(M_\mathrm{QCD}\right)^{b}_{a} = \left(\bar{q}_R\right)^{b}\left(q_L\right)_a \Rightarrow \left(S_\mathrm{QCD}\right)^{b}_{a} = q_{a}(x) \bar{q}^b(x),
\end{equation}
where $\{a,b\}$ are flavor indices. Four-quark operators may also be used to construct a quark-level representation of $M'_\mathrm{QCD}$. In this way, the chiral nonets may be expressed in a complementary way by means of the constituent quark fields, and ultimately QCD composite operators,
\begin{gather}
    M_\mathrm{QCD} = S_\mathrm{QCD}+ i\phi_\mathrm{QCD},\quad {M'}_\mathrm{QCD} = {S'}_\mathrm{QCD} + i{\phi'}_\mathrm{QCD},
\end{gather}
where $S_\mathrm{QCD}$ and ${S'}_\mathrm{QCD}$ written in terms of QCD composite operators (currents) are
\begin{equation}
    \!S_{\rm QCD}=
\begin{pmatrix}
J_1^{11} & J_1^{a_0^+} & J_1^{\kappa^+}  \\
J_1^{a_0^-} & J_1^{22} & J_1^{\kappa^0} \\
J_1^{\kappa^-} & J_1^{{\bar \kappa}^0} & J_1^{33}
\end{pmatrix}, 
~
S'_{\rm QCD} =
\begin{pmatrix}
J_2^{11} & J_2^{{a'}_0^+} & J_2^{{\kappa'}^+}  \\
J_2^{{a'}_0^-} & J_2^{22} & J_2^{{\kappa'}^0} \\
J_2^{{\kappa'}^-} & J_2^{{\bar {\kappa'}}^0} & J_2^{33} \\
\end{pmatrix},
\label{eq:QCD-scalarnonet}
\end{equation}
where $J_1$ represents a two-quark current and $J_2$ represents a four-quark current. After representing the chiral nonets both in terms of effective fields (within the framework of chiral Lagrangians) and in terms of constituent quark fields (within the framework of QCD sum-rules), we connect these representations by introducing scale factor matrices $\{I_M,I_{M'}\}$ constrained by the chiral symmetry transformations in eq.~\eqref{eq:chiral-symmetry-trans},
\begin{gather}
    M = I_M M_\mathrm{QCD},\quad {M'} = I_{M'} {M'}_\mathrm{QCD},
\end{gather}
where the scale factor matrices have the form 
\begin{equation}
    I_M = -\frac{m_q}{\Lambda^3}\times \mathds{1},\quad I_{M'}= \frac{1}{{\Lambda'}^5}\times \mathds{1}.
    \label{eq:universality}
\end{equation}
Here, the quark mass $m_q = (m_u+m_d)/2$ has the appropriate renormalization group behavior necessary for the QCD currents. The constants $\{\Lambda,{\Lambda'}\}$ are energy-independent scale factors which have dimensions of energy, and define the property of \textit{universality}; the scale factor parameters $\{\Lambda,{\Lambda'}\}$ must be identical for all members of the nonet. We have previously confirmed universality exhibited in the isotriplet $a_0$ and isodoublet $K_0^{*}$ sectors \cite{CLQCDSR_2016,CLQCDSR_2020}, and this work builds a foundation for extending our analysis to the isoscalar sector.

The effective and constituent-level representations of the scalar nonet must ultimately be related to the physical mesonic fields. The physical mesonic fields $\mathbf{H} = \{\mathbf{K},\mathbf{A}\}$ are expected to be a mixture of two-quark, four-quark, and gluonic fields \cite{CLQCDSR_2016,CLQCDSR:2019_Proc,CLQCDSR_2020}, and can be related to the chiral Lagrangian effective fields through a rotation matrix $L_s$ dependent upon a mixing angle between the physical and chiral Lagrangian fields $\theta_s$, where the subscript $s = \{\kappa,a\}$ indicates the isospin of the system. For the isodoublet system relating the physical states $K_0^{*}\left(700\right)$ and $K_0^{*}\left(1430\right)$ to the appropriate chiral Lagrangian fields and QCD currents,
\begin{equation}
    \mathbf{K} = \begin{pmatrix}
K_0^*(700)\\
K_0^*(1430)
\end{pmatrix}
= L_\kappa^{-1}
\begin{pmatrix}
S^3_2\\
\left(S'\right)^3_2
\end{pmatrix}
=
{L_\kappa^{-1} I_\kappa J_\kappa^{\rm QCD} },
\label{K_def_new}
\end{equation}
where the scale factor and rotation matrices are given by 
\begin{equation}
    L^{-1}_\kappa=\begin{pmatrix}
\cos\theta_\kappa & -\sin\theta_\kappa
\\
\sin\theta_\kappa & \cos\theta_\kappa
\end{pmatrix}
\,,~I_\kappa =
\begin{pmatrix}
\frac{-m_q}{\Lambda^3} &0 \\
0 & \frac{1}{{{\Lambda'}^5}}
\end{pmatrix}
~,
\end{equation}
and the QCD currents are
\begin{gather}
    J_\kappa^{\rm QCD}=\begin{pmatrix}
J^{\kappa}_1\\
J^{\kappa}_2
\end{pmatrix}
\,,~ J^{\kappa}_1=\bar ds 
\label{J1_kappa}
\\[5pt]
    J^{\kappa}_2=\sin(\phi) u^T_\alpha C\gamma_\mu\gamma_5 s_\beta\left(\bar d_\alpha\gamma^\mu\gamma_5 C\bar u_\beta^T-\alpha\leftrightarrow \beta \right)
+\cos(\phi) d^T_\alpha C\gamma_\mu s_\beta\left(\bar d_\alpha\gamma^\mu C\bar u_\beta^T+\alpha\leftrightarrow \beta \right)
\label{J2_kappa}
\end{gather}
where $C$ is the charge conjugation operator and $\cot\phi=1/\sqrt{2}$ \cite{Chen:2007xr}. Likewise, for the isotriplet system relating the physical states $a_0\left(980\right)$ and $a_0\left(1450\right)$,
\begin{equation}
    \mathbf{A} = \begin{pmatrix}
a_0^0(980)\\
a_0^0(1450)
\end{pmatrix}
= L_a^{-1}
\begin{pmatrix}
\frac{S_1^1 - S_2^2}{\sqrt{2}}\\
\frac{{S'}_1^1 - {S'}_2^2}{\sqrt{2}}
\end{pmatrix}
=
L_a^{-1} I_a J_a^{\rm QCD}.
\label{A_def_new}
\end{equation}
where the scale factor and rotation matrices are given by
\begin{equation}
    L^{-1}_\kappa=\begin{pmatrix}
\cos\theta_\kappa & -\sin\theta_\kappa
\\
\sin\theta_\kappa & \cos\theta_\kappa
\end{pmatrix}
\,,~I_\kappa =
\begin{pmatrix}
\frac{-m_q}{\Lambda^3} &0 \\
0 & \frac{1}{{{\Lambda'}^5}}
\end{pmatrix}
~,
\label{I_matrices_new}
\end{equation}
and the QCD currents are
\begin{gather}
    J_a^{\rm QCD} 
=\begin{pmatrix}
J^a_1\\
J^a_2
\end{pmatrix}\,,~
J^a_1=\left(\bar u u-\bar d d\right)/\sqrt{2}
\label{J_a_QCD_new}
\\[5pt]
J^a_2=\frac{\sin\phi}{\sqrt{2}}d^T_\alpha C\gamma_\mu\gamma_5 s_\beta\left(\bar d_\alpha\gamma^\mu\gamma_5 C\bar s_\beta^T-\alpha\leftrightarrow \beta \right)
+\frac{\cos\phi}{\sqrt{2}}d^T_\alpha C\gamma_\mu s_\beta\left(\bar d_\alpha\gamma^\mu C\bar s_\beta^T+\alpha\leftrightarrow \beta \right)
- u\leftrightarrow d\,,
\label{J2_a}
\end{gather}

\subsection{Revising the scale factor methodology}
We consider a slight revision of the original scale factor methodology presented in \cite{CLQCDSR_2016,CLQCDSR:2019_Proc,CLQCDSR_2020} (a detailed treatment may be found in \cite{ExtBridge}). Previously, an off-diagonal constraint was implemented in order to relate Gaussian sum-rules (GSRs) \cite{gauss,harnett_quark} to the physical correlator. However to generalize our methodology to higher-dimensional isospin systems, a different methodological choice will simplify the analysis. Rearranging eqs.~\eqref{K_def_new} and \eqref{A_def_new}, and placing the rotation matrix $L_s$ with the physical states allows us to relate the QCD currents directly to mixtures of the physical chiral Lagrangian hadronic fields,
\begin{gather}
   L_\kappa\mathbf{K} =  L_\kappa\begin{pmatrix}
K_0^*(700)\\
K_0^*(1430)
\end{pmatrix}
= 
\begin{pmatrix}
S^3_2\\
\left(S'\right)^3_2
\end{pmatrix}
=
{ I_\kappa J_\kappa^{\rm QCD} },\\
L_a\mathbf{A} = L_a\begin{pmatrix}
a_0^0(980)\\
a_0^0(1450)
\end{pmatrix}
= 
\begin{pmatrix}
\frac{S_1^1 - S_2^2}{\sqrt{2}}\\
\frac{{S'}_1^1 - {S'}_2^2}{\sqrt{2}}
\end{pmatrix}
=
 I_a J_a^{\rm QCD}.
\end{gather}
From here, correlation functions can be constructed. The physical hadronic fields represented by $\mathbf{H} = \{\mathbf{K},\mathbf{A}\}$ form diagonal correlation functions,
\begin{gather}
\Pi^{\rm H}_{ij}  \left(Q^2 =-q^2\right)
=i\int d^4x\, e^{iq\cdot x}
\langle 0| {\rm T} \left[ {\bf H}_i (x) {\bf H}_j(0) \right] |0 \rangle\,,
\label{had_diag_property}
\end{gather} 
where $\Pi^{\rm H}_{ij} = 0$ for $i\ne j\,$, so forming diagonal QCD correlation functions results in the following relationships:
\begin{gather}
    \frac{m_q^2}{\Lambda^6}\Pi^{\rm QCD}_{11}\left(Q^2\right) =\cos^2{\theta_s}\Pi^H_{11}\left(Q^2\right)+  \sin^2{\theta_s}\Pi^H_{22}\left(Q^2\right),\,
 \label{Pi_QCD_11}
 \\
 \frac{1}{\left(\Lambda'\right)^{10}}\Pi^{\rm QCD}_{22}\left(Q^2\right) =\sin^2{\theta_s}\Pi^H_{11}\left(Q^2\right)+  \cos^2{\theta_s}\Pi^H_{22}\left(Q^2\right)\,,
 \label{Pi_QCD_22}
\end{gather}
where the QCD correlation functions are defined according to
\begin{equation}
    \Pi^{\rm QCD}_{mn}\left(Q^2=-q^2\right)= i\int d^4x\, e^{iq\cdot x}\langle 0| {\rm T}  \left[ J^{\rm QCD}_m (x) J_n^{\rm QCD}(0)^\dagger \right] |0 \rangle.
\end{equation}
From here, we apply a GSR methodology to these correlation functions \cite{gauss,harnett_quark}, which parametrizes the hadronic spectral function into a resonance contribution $\trho^{\rm res}(t)$ and a continuum contribution characterized by a continuum threshold parameter $s_0$
\begin{equation}
\trho^H(t)=
\trho^{\rm res}(t) +\theta\left(t-s_0\right){\frac{1}{\pi}} {\rm Im} \Pi^{ \rm QCD}(t)\,.
\label{spectral_new}
\end{equation}
Applying a GSR methodology to eqs.~\eqref{Pi_QCD_11} and \eqref{Pi_QCD_22} results in 
\begin{gather}
    \frac{m_q^2}{\Lambda^6}{\cal G}^{\rm QCD}_{(s)11}\left({\hat s}, \tau, s_0^{(1)}\right) =\cos^2{\theta_s}G_{(s)11}^{\rm res} \left({\hat s}, \tau\right)+  \sin^2{\theta_s}G_{(s)22}^{\rm res} \left({\hat s}, \tau\right)\,,
 \label{GSR_11}
 \\
 \frac{1}{\left(\Lambda'\right)^{10}}{\cal G}^{\rm QCD}_{(s)22}\left({\hat s}, \tau, s_0^{(2)}\right) =\sin^2{\theta_s}G_{(s)11}^{\rm res} \left({\hat s}, \tau\right)+  \cos^2{\theta_s}G_{(s)22}^{\rm res} \left({\hat s}, \tau\right)\,,
 \label{GSR_22}
\end{gather}
where the subscripts are inherited from the corresponding correlation functions in eqs.~\eqref{Pi_QCD_11} and \eqref{Pi_QCD_22}, and the GSR expressions associated with the hadronic resonance pieces and the QCD description are given by
\begin{gather}
    G^{\rm res} \left({\hat s}, \tau\right) =\frac{1}{\sqrt{4\pi\tau}}
\int\limits_{t_0}^{\infty} \!\! dt \,{\rm exp} \left[  {\frac{-({\hat s} - t)^2}{4\tau}}\right]\,\trho^{\rm res} (t)\,,  
\label{gsr_res}
\\
{\cal G}^{\rm QCD}\left({\hat s}, \tau, s_0\right)=
G^{\rm QCD} \left({\hat s}, \tau\right)-
\frac{1}{\sqrt{4\pi\tau}}
\int\limits_{s_0}^{\infty} \!\! dt \,{\rm exp} \left[  {\frac{-({\hat s} - t)^2}{4\tau}}\right]\,\frac{1}{\pi}{\rm Im}\Pi^{\rm QCD}(t)\,.  
\label{gsr_QCD_cont}
\end{gather}
The mixing angles $\theta_s$ that appear in eqs.~\eqref{GSR_11} and \eqref{GSR_22} are determined from a chiral Lagrangian analysis \cite{GLSM} to be $\cos\theta_\kappa=0.4161$ and $\cos\theta_a=0.6304$. Eqs.~\eqref{GSR_11} and \eqref{GSR_22} may then be rewritten as
\begin{gather}
    G_{(s)11}^{\rm res}(\hat s,\tau)=a A_s
{\cal G}_{(s)11}^{\rm QCD}\left(\hat s,\tau, s_0^{(1)}\right)-bB_s
{\cal G}_{(s)22}^{\rm QCD}\left(\hat s,\tau, s_0^{(2)}\right),
\label{G_eqs_1}
\\
G_{(s)22}^{\rm res}(\hat s,\tau)=-aB_s
{\cal G}_{(s)11}^{\rm QCD}\left(\hat s,\tau, s_0^{(1)}\right)+bA_s
{\cal G}_{(s)22}^{\rm QCD}\left(\hat s,\tau, s_0^{(2)}\right),
\label{G_eqs_2}
\end{gather}
where the coefficients $\{A_s,B_s\}$ and $\{a,b\}$ are
\begin{gather}
    A_s=\frac{\cos^2\theta_s}{\cos^2\theta_s-\sin^2\theta_s}\,,~
B_s=\frac{\sin^2\theta_s}{\cos^2\theta_s-\sin^2\theta_s}\,,\quad
a=\frac{m_q^2}{\Lambda^6}\,,~b=\frac{1}{\left(\Lambda'\right)^{10}}\,.
\end{gather}
\section{Modeling resonance shapes using chiral Lagrangians}
In our previous work, we modeled the physical hadronic resonances $\trho^{\rm res}_i$ in eq.~\eqref{gsr_res} with a Breit-Wigner resonance shape, 
\begin{gather}
\trho_i^{\rm prop} (t)  ={\rm Im}\Pi^H_{ii}(t)=\frac{m_i\Gamma_i}{\left(t-m_i^2\right)^2+m_i^2\Gamma_i^2}
\,,
\label{eq.rho_prop}
\end{gather}
which is a generalization of the typical narrow resonance (NR) shape often used in GSR analyses, recovered from \eqref{eq.rho_prop} in the $\Gamma_i\rightarrow 0$ limit, $\trho_i^{\rm NR} (t) =\delta\left(t-m_i^2\right)$.
However, given the nature of wide resonances such as the $K_0^{*}(700)$, we seek to develop more sophisticated resonance models inspired by the chiral Lagrangian methodology. 
\subsection{The kappa system}
\begin{figure}[htb]
	\centering
	\includegraphics[width=0.31\columnwidth]{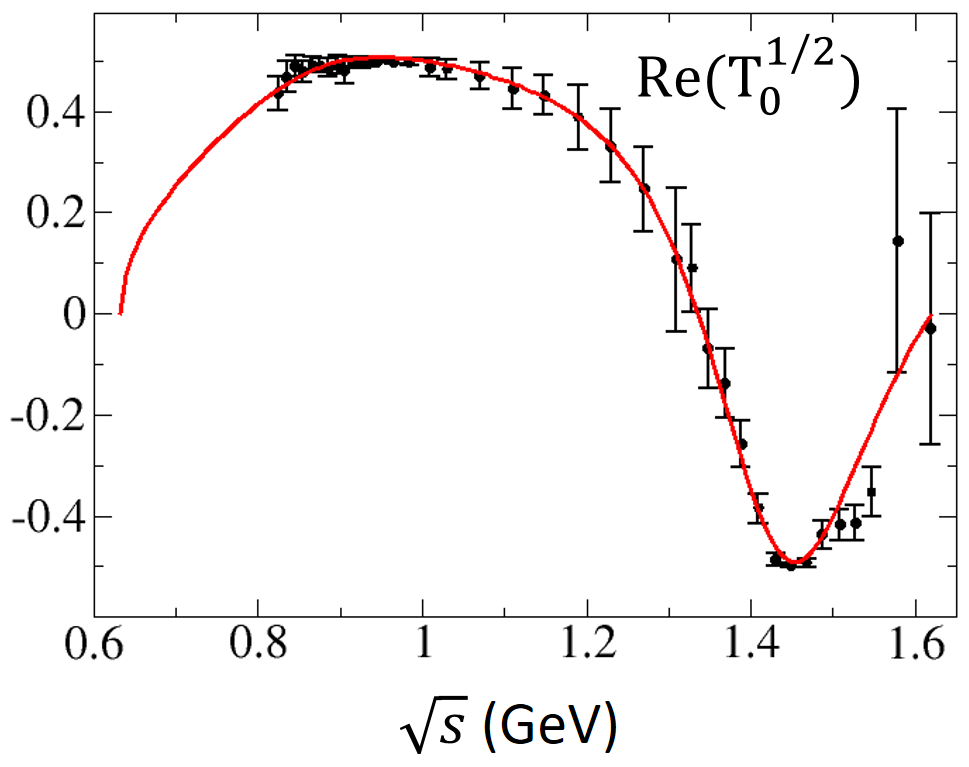}\hspace{0.01\columnwidth}
    \includegraphics[width=0.31\columnwidth]{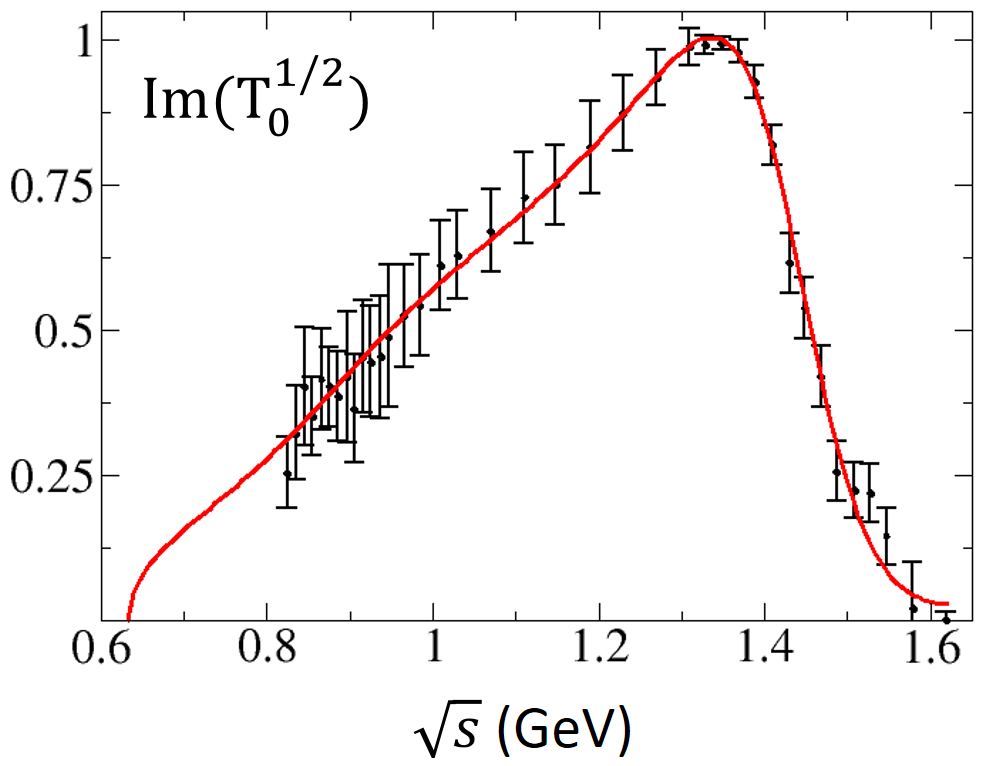}\hspace{0.01\columnwidth}
    \includegraphics[width=0.31\columnwidth]{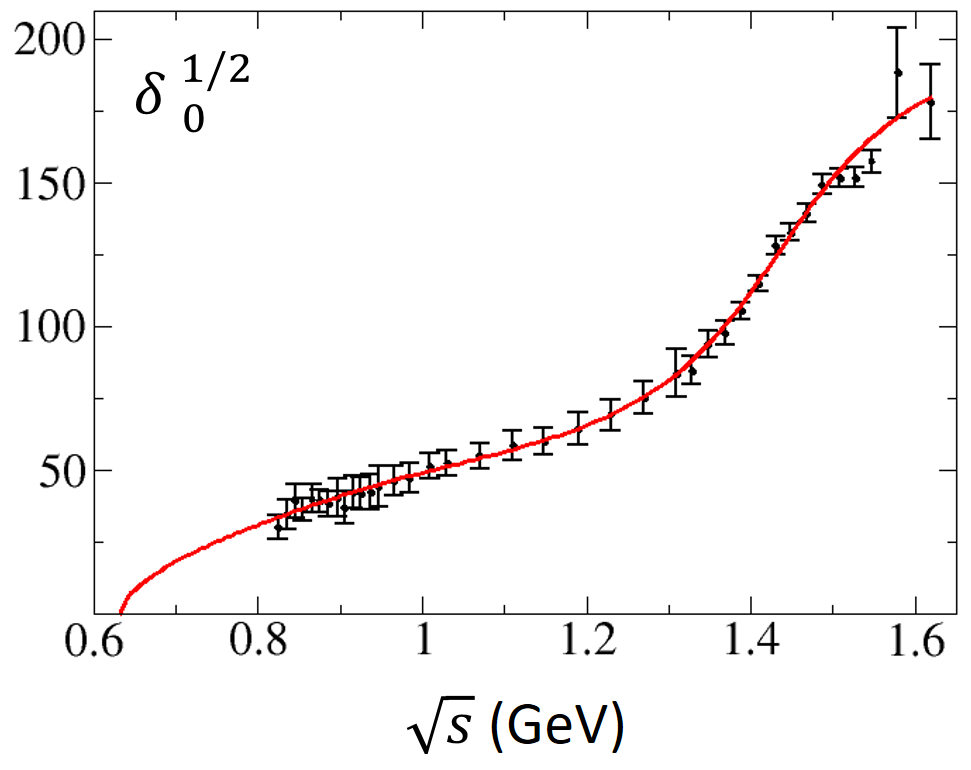}
\caption{Fit of the background-resonance interference approximation model (\ref{E_T012_rBW})   
(solid red lines),  
to the experimental data \cite{Aston} (solid dots and error bars).
The first two figures show the real and imaginary parts of the $I=1/2$, $J=0$, $\pi K$  scattering amplitude, followed by the third figure that shows the phase shift.  }
 \label{F_rBW}
\end{figure}

Previous work studying the $\pi K$ scattering amplitude within the framework of a generalized linear sigma model \cite{GLSM_piK} has shown that the unitarized scattering amplitude can be modeled using a $K$-matrix methodology, approximating the isospin $I=1/2$ scalar scattering amplitude by two poles and a background,
\begin{equation}
T_0^{1/2}(s) = \frac{\rho(s)}{2} 
\left[
C_\kappa +
\sum_{i=1}^2\,
 {
	 \frac{A_{\kappa_i}\, \left( 2 m_{\kappa_i}\, \Gamma_{\kappa_i}\right)}
  { \rho_{0i} \left(m_{\kappa_i}^2 - s\right)  - i\, m_{\kappa_i} \Gamma_{\kappa_i} \rho_{0i}}
 }
\right],
\label{E_T012_rBW}
\end{equation}
where 
\begin{equation}
\rho(s) = \frac{q} 
{8 \pi \sqrt{s}}
= 
\frac{\sqrt{
	[s-(m_\pi+m_K)^2]
	[s-(m_\pi-m_K)^2]
}} 
{16 \pi s},
\label{E_rho_piK}
\end{equation} 
with $q$ representing the center of mass momentum, and where $\rho_{0i} = \rho(m_{\kappa_i}^2)$,  with $i=1, 2$ representing the light and the heavy kappas, respectively. In this work, we introduce the \textit{background-resonance interference approximation}, extending the model of \cite{GLSM_piK} to include complex coefficients, taking the constants to be $A_{\kappa_i}= A_{\kappa_iR}+i A_{\kappa_iI}$ and $C_{\kappa}= C_{\kappa R}+i C_{\kappa I}$.  
These constants are then determined from fits to the available experimental data on $\pi K$ scattering \cite{Aston}. By extending the model to contain complex coefficients, information about the interference between the two resonances may be captured. Additionally, both the real and imaginary parts of the resonances will contribute to the real and imaginary parts of the scattering amplitude. We find that this generalization of the model introduced in \cite{GLSM_piK} demonstrates notable improvement from the original work; the results of a combined unweighted $\chi^2$ fit of the parameters $\{A_{\kappa_1R}, A_{\kappa_1I}, A_{\kappa_2 R}, A_{\kappa_2 I}, C_{\kappa R}, C_{\kappa I} \}$ are summarized in Table \ref{T_BW_EBW_fits}, while a comparison against the available experimental data is shown in Figure \ref{F_rBW}. We constructed a second parameterization of the scattering amplitude \cite{ExtBridge}, modifying each resonance contribution to the scattering amplitude to be unitary through the addition of an energy-dependent imaginary part,
\begin{equation}
T_0^{1/2} (s) = \frac{\rho (s)} 
{2} 
\left[
C_\kappa +
\sum_{i=1}^2\,
\frac{
A_{\kappa_i}\, 	\left(2 m_{\kappa_i}  \Gamma_{\kappa_i} \right)}
	{\rho_{0i} \, \left( m_{\kappa_i}^2 - s \right) - i\, m_{\kappa_i} \Gamma_{\kappa_i} \, \rho(s)}
\right]\,.
\label{E_T012_rEBW}
\end{equation}
The resulting fitting coefficients for this model are also shown in the Table \ref{T_BW_EBW_fits}, along with a goodness of fit measure $\Delta\delta_{0}^{1/2}$ calculating the average relative error between model computation of phase shift and the central values of the experimental data on phase shift
\begin{equation}
    \Delta\delta_{0}^{1/2} = \frac{1}{N}\, 
\sum_k^N 
{ 
	\frac{\left| \delta_0^{1/2, {\rm Theo.} } (s_k)- \delta_0^{1/2, {\rm Exp.}} (s_k)  \right|}{\delta_0^{1/2, {\rm Exp.}} (s_k)}.
}
\label{E_pik_fit_goodness}
\end{equation}
The fits between model (\ref{E_T012_rBW}) and model (\ref{E_T012_rEBW}) have negligibly small deviations; plots demonstrating the comparison between model (\ref{E_T012_rEBW}) and the experimental data have been omitted and may be found in our original work \cite{ExtBridge}.
\begin{table}[ht]
\renewcommand{\arraystretch}{1.0}
\centering
	\begin{tabular}{c|c|c|c||c}
		\hline
		\rule{0pt}{3ex}   
	 & $A_{\kappa_1}$ &  $A_{\kappa_2}$ & $C_\kappa$  & $\Delta \delta_0^{1/2}$
		\\[2pt]
		\hline
		Model (\ref{E_T012_rBW})	 &  $-0.0560 - i\, 0.144$    &   $0.385 + i\,  0.899$ & $47.853 + i\,  33.116$ & 0.026
		\\
		Model (\ref{E_T012_rEBW})   &  $-0.0559 - i\,  0.150$ &  $0.382 + i\, 0.900$ & $46.861 + i\, 33.386 $ & 0.028
	\end{tabular}
	\caption{The  complex parameters of the background-resonance interference approximation models (\ref{E_T01_rBW}) and (\ref{E_T01_a0_EBW}) determined from fits to theoretical prediction of \cite{00_BFS61}.}
	\label{T_BW_EBW_fits}
\end{table}
\subsection{The $a_0$ system}
\begin{figure}[htb]
	\centering
	\includegraphics[width=0.31\columnwidth]{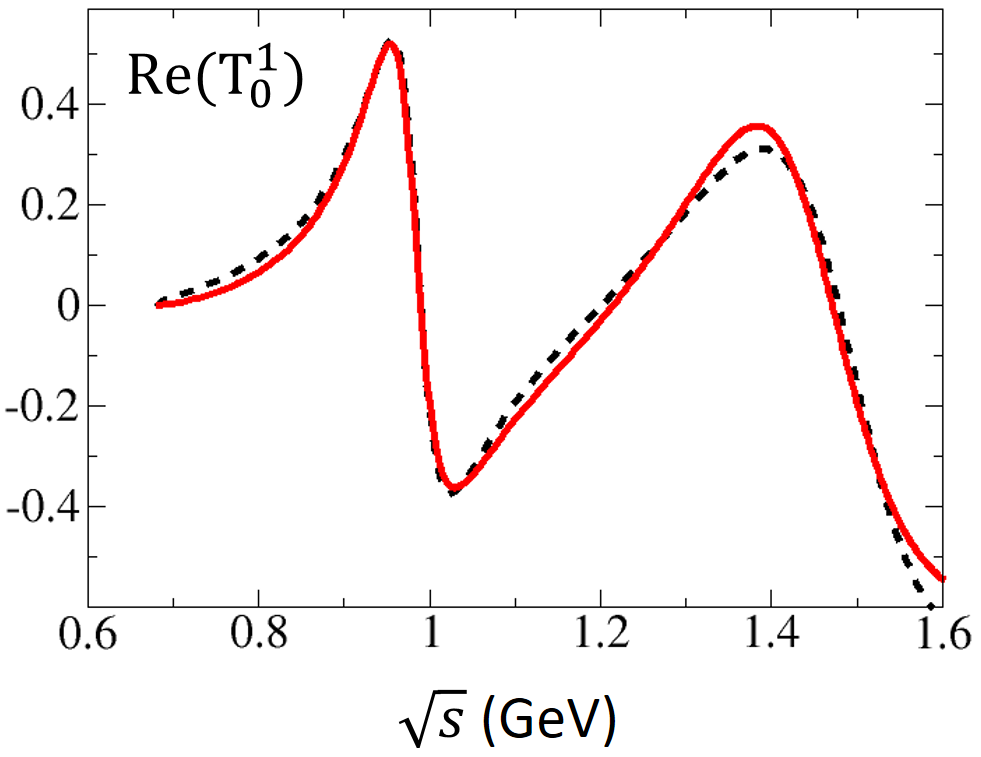}\hspace{0.01\columnwidth}
    \includegraphics[width=0.31\columnwidth]{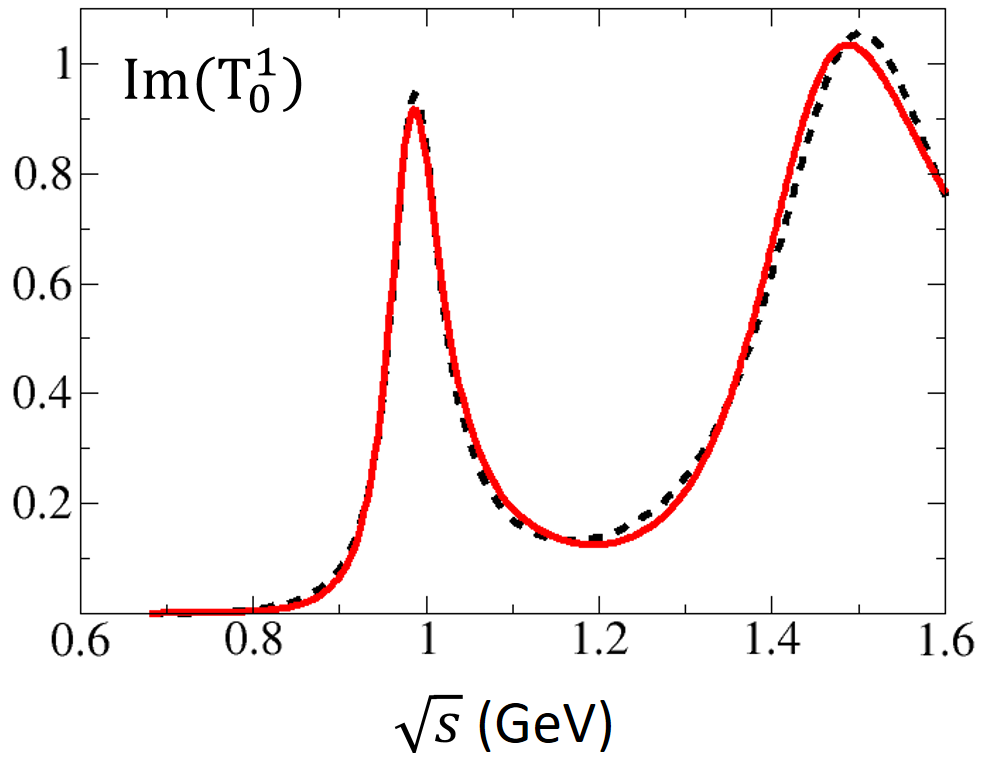}\hspace{0.01\columnwidth}
    \includegraphics[width=0.31\columnwidth]
{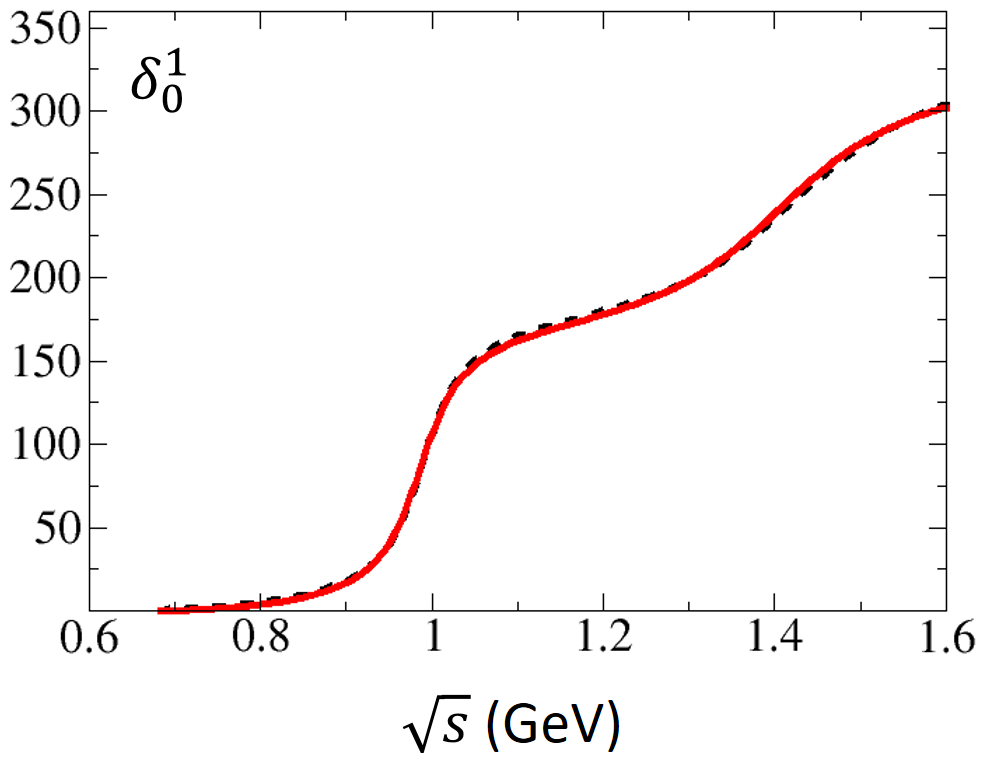}
\caption{
Fit of the background-resonance interference approximation model (\ref{E_T01_rBW}) (solid red lines),  
to the theoretical predictions of \cite{00_BFS61} (dashed  lines).
The first two figures show the real and imaginary parts of the $I=1$, $J=0$, $\pi \eta$  scattering amplitude, followed by the third figure that shows the phase shift.  
	}
	\label{F_pieta_fit_BW}
\end{figure}

In a similar manner, we construct a model for the isosinglet scalar mesons ($I = 1, \, J=0$) within the same background-resonance interference approximation. Considering the tree-level $\pi\eta$ scattering amplitude as a first approximation to the $a_0$ shape, we parameterize a two-pole plus background unitarized scattering amplitude as
\begin{equation}
T_0^1(s) = \frac{\rho(s)}{2} 
\left[
C_a +
\sum_{i=1}^2\,
 {
	 \frac{A_{a_i}\, \left( 2 m_{a_i}\, \Gamma_{a_i}\right)}
  { \rho_{0i} \left(m_{a_i}^2 - s\right)  - i\, m_{a_i} \Gamma_{a_i} \rho_{0i}}
 }
\right],
\label{E_T01_rBW}
\end{equation}
where 
\begin{equation}
\rho(s) = \frac{q}
{8 \pi \sqrt{s}}
= 
\frac{\sqrt{
	\left[s - (m_\pi + m_\eta)^2\right]
	\left[s-  (m_\pi - m_\eta)^2\right] 
}}
{16 \pi s},
\label{E_rho_pieta}
\end{equation} 
with $q$ the center of mass momentum, 
 $\rho_{0i} = \rho(m_{a_i}^2)$,  with $i=1, 2$ representing the light and the heavy $a_0$. As previously, the complex constants $C_a = C_{aR} + i\, C_{aI}$, $A_{a_i} = A_{a_i R} + i\, A_{a_i I}$ have to be determined from fits to the scattering amplitude. No direct experimental data on $\pi\eta$ scattering is available, and so we fit these constants to a theoretical prediction for this scattering amplitude  \cite{00_BFS61}. Also following the previous analysis of the kappa system, a second parameterization with local unitarization is considered,
 \begin{equation}
T_0^1(s) = \frac{\rho(s)}{2} 
\left[
C_a +
\sum_{i=1}^2\,
	 \frac{A_{a_i}\, \left( 2 m_{a_i}\, \Gamma_{a_i}\right)}
  { \rho_{0i} \left(m_{a_i}^2 - s\right)  - i\, m_{a_i} \Gamma_{a_i} \rho(s)}
\right].
\label{E_T01_a0_EBW}
\end{equation}
 The result of the same combined unweighted $\chi^2$ fitting procedure applied to the kappa system results in the fitting coefficients summarized in Table \ref{T_BW_EBW_a0_fits}, with Figure~\ref{F_pieta_fit_BW} demonstrating the agreement between model \eqref{E_T01_rBW} and calculations from \cite{00_BFS61}. Again, there are negligible differences between model \eqref{E_T01_rBW} and \eqref{E_T01_a0_EBW}, and so only the comparisons between model \eqref{E_T01_rBW} and the data are shown here; further detail may be found in \cite{ExtBridge}.
\begin{table}[ht]
\renewcommand{\arraystretch}{1.0}
\centering
	\begin{tabular}{c|c|c|c}
		\hline
		\rule{0pt}{3ex}   
	 & $A_{a_1}$ &  $A_{a_2}$ & $C_a$   
		\\[2pt]
		\hline
		Model (\ref{E_T01_rBW})	 &  $0.864 - i\, 0.298$    &   $0.726 - i\,  0.641$ & $-42.7 + i\,  17.6$
		\\
		Model (\ref{E_T01_a0_EBW})   &  $0.872 - i\,  0.288$ &  $0.710 - i\, 0.635$ & $-43.0 + i\, 19.9$
	\end{tabular}
	\caption{The  complex parameters of the background-resonance interference approximation models (\ref{E_T01_rBW}) and (\ref{E_T01_a0_EBW}) determined from fits to theoretical prediction of \cite{00_BFS61}.}
	\label{T_BW_EBW_a0_fits}
\end{table}

\section{Gaussian sum-rules with enhanced resonance models}
Considering the modeling of the scattering amplitudes with a unitarized $K$-matrix methodology, we define a new series of resonance models to compare against our previous GSR analysis using a Breit-Wigner, or ``propagator'' (PROP) resonance model \cite{CLQCDSR_2020}. The \textbf{extended distorted Breit-Wigner} model (EDBW) is constructed by adding a factor of $\rho(t)$ and a factor $\xi_i$ describing the relative contributions of the real and imaginary parts of the complex resonance poles,
\begin{gather}
 \trho^{\rm EDBW}_i(t)  =\frac{\rho(t)m_i\Gamma_i}{\left(t-m_i^2\right)^2+m_i^2\Gamma_i^2}+\xi_i \frac{\rho(t)\left(m_i^2-t\right)}{\left(t-m_i^2\right)^2+m_i^2\Gamma_i^2}\,.
    \label{rho_xi_fixed_width_new}
\end{gather}
Two values of $\xi_i$ are possible given our choice of normalization,  
\begin{equation}
    \xi_i^{(1)} =  - \frac{A_{iR}}{A_{iI}}\,,\quad \xi_i^{(2)} =  \frac{A_{iI}}{A_{iR}}.
    \label{xi}
\end{equation}
Examining the real parts of the chiral Lagrangian models \eqref{E_T012_rBW} and \eqref{E_T01_rBW} gives $\xi_i^{(1)}$, while examining the imaginary parts of the same chiral Lagrangian model gives $\xi_i^{(2)}$. Both solutions are considered in the analysis, however using the requirement of GSR positivity we can eliminate one of the possible values of $\xi_i$ (for more detail, see \cite{ExtBridge}). 

By examining our second local unitary parameterization of the scattering amplitude, we can generate similar resonance models corresponding to our chiral Lagrangian models \eqref{E_T012_rEBW} and \eqref{E_T01_a0_EBW}. The \textbf{extended generalized Breit-Wigner} model (EGBW) is defined as,
\begin{gather}
\trho^{\rm EGBW}_i(t)  =\frac{m_i G_i(t)^2}{\left(t-m_i^2\right)^2+m_i^2G_i(t)^2}+\xi_i \frac{G_i(t)\left(m_i^2-t\right)}{\left(t-m_i^2\right)^2+m_i^2G_i(t)^2}\,,\quad
G_i(t)=\frac{\rho(t)\Gamma_i}{\rho_{0i}}\,, \label{rho_xi_energy_width_new}
\end{gather}
where the expressions for $\xi_i$ in eq.~\eqref{xi} also apply here. In the case of both the EDBW and EGBW models, a cut-off is implemented to improve convergence \cite{ExtBridge}. As well, both the EDBW and EGBW models reduce to different models in the case of $\xi_i = 0$; in the case of the EDBW model, $\xi_i = 0$ gives what we will call the \textbf{distorted Breit-Wigner} model (DBW), and in the case of the EGBW model, we obtain what we will call the \textbf{generalized Breit-Wigner} model (GBW). Including the PROP model from our earlier analysis, we form a set of models with increasing sophistication to evaluate within a GSR analysis. In order to consistently compare against our previous work \cite{CLQCDSR_2016, CLQCDSR_2020}, we normalize each resonance to the PROP model in the revised methodology within the GSR energy range $0 < \hat{s} <6\,\mathrm{GeV}^2$. We use the parameters summarized in Table \ref{gsr_res_parameter_tab} in our GSR analysis.
\begin{table}[!h]
\centering
\renewcommand{\arraystretch}{0.8}
\begin{tabular}{|c|c|c|c|c|c|c|}
\hline
Resonance & Mass (GeV) & Width (GeV) & $\xi^{(1)}$ (EDBW) & $\xi^{(2)}$ (EDBW) & $\xi^{(1)}$ (EGBW)   & $\xi^{(2)}$ (EGBW)
\\
\hline
$K_0^*(700)$ & $0.824$ & $0.478$ & $-0.3889$ & $2.571$ 
& $-0.4283$ & $2.335$
\\
\hline
$K_0^*(1430)$ & $1.425$ & $0.270$ & $-0.3727$ & $2.683$ 
& $-0.4244$ & $2.356$
\\
\hline
$a_0(980)$ & $0.980$ & $0.06$ &  $2.899$  &   $-0.3449$ 
&  $3.027$  &  $-0.3303$  
\\
\hline
$a_0(1450)$ & $1.47$ & $0.265$ &  $1.132$  &   $-0.8835$ 
&  $1.119$  &  $-0.8937$ 
\\
\hline
\end{tabular}
\caption{
Parameters used for the resonance models.  The parameters $\{\xi^{(1)}_i,\xi^{(2)}_i\}$ are obtained from 
(\ref{xi}) combined with Tables~\ref{T_BW_EBW_fits} and \ref{T_BW_EBW_a0_fits}. The DBW and GBW models are evaluated with $\xi_i = 0$. The masses and widths correspond to our benchmark analysis \cite{CLQCDSR_2020} and are consistent with the PDG ranges \cite{PDG}.}
\label{gsr_res_parameter_tab}
\end{table}

\section{Results}
We use available correlation functions associated with the two- and four-quark currents \eqref{J1_kappa}, \eqref{J2_kappa}, \eqref{J_a_QCD_new} and \eqref{J2_a} \cite{Zhang:2009qb,Du:2004ki,Jamin:1992se,Reinders:1984sr,Chen:2007xr}, and apply the methods of \cite{harnett_quark}. The QCD input parameters used are \cite{Reinders:1984sr,Narison:2011rn,Beneke:1992ba,Belyaev:1982sa}
\begin{gather}
 \langle\alpha_s G^2\rangle
      = (0.07\pm 0.02)\, {\rm GeV^4} \,,
      \label{GG}
      \\
\frac{\left\langle \overline{q}\sigma G q\right\rangle}{\langle \bar q q\rangle}=\frac{\left\langle \overline{s}\sigma G s\right\rangle}{\langle \bar s s\rangle}=(0.8\pm 0.1) \,{\rm GeV^2}
\label{mix}
\\
\langle \bar q q\rangle=-\left(0.24\pm 0.2 \,{\rm GeV}\right)^3\,,~\langle \bar s s\rangle=(0.8\pm 0.1)\langle \bar q q\rangle
\label{O6}
\\
2\mqq=-f_\pi^2m_\pi^2\,, f_\pi = 130/\sqrt{2} {\rm MeV}
\\
  n_{{c}} = 8.0\times 10^{-4}\ {\rm GeV^4}~,~\rho_c =1/600\,{\rm MeV}~,
  \label{inst}
  \\
  m^*_q=170\,{\rm MeV}\,,~m^*_s=220\,{\rm MeV}~.
\end{gather}
Values for quark masses, quark mass ratios, and $\alpha_s$ are taken from the Particle Data Guide (PDG) \cite{PDG}. As in our benchmark scale factor analysis \cite{CLQCDSR_2020} and as in other GSR analyses \cite{harnett_quark}, we define our renormalization scale at $\nu^2=\sqrt{\tau}$ with a value of $\tau=3\,{\rm GeV^4}$. Vacuum saturation has been used  for the dimension-six (four-quark) condensates. From eqs.~\eqref{G_eqs_1} and \eqref{G_eqs_2}, expressions for the scale factors $\{\Lambda,\,\Lambda'\}$ can be written as
\begin{gather}
\Lambda=\left[\frac{\left(A_s+B_s\right) m_q^2 \,{\cal G}_{(s)11}^{\rm QCD}\left(\hat s,\tau, s_0^{(1)}\right) }{A_s G_{(s)11}^{\rm res}(\hat s,\tau)+B_sG_{(s)22}^{\rm res}(\hat s,\tau)}\right]^{1/6}\equiv \lambda_s\left(\hat s,\tau,s_0^{(1)}\right)
\,,
\label{scale_relation_lambda}
\\[2pt]
\Lambda'=\left[\frac{\left(A_s+B_s\right) \, {\cal G}_{(s)22}^{\rm QCD}\left(\hat s,\tau, s_0^{(2)}\right) }{B_s G_{(s)11}^{\rm res}(\hat s,\tau)+A_sG_{(s)22}^{\rm res}(\hat s,\tau)}\right]^{1/10} \equiv \lambda'_s\left(\hat s,\tau,s_0^{(2)}\right)
\,,
\label{scale_relation_lambda_prime}
\end{gather}
combining a factor of $m_q^2$ into ${\cal G}_{(s)11}^{\rm QCD}$ via renormalization-group improvement \cite{ExtBridge}. The scale factors are fitted to $\{\lambda_s,\lambda'_s\}$ over equally-spaced $\hat s$ values (usually about 25 points) \cite{CLQCDSR_2020,ExtBridge}, and continuum parameters $\{s_0^{(1)},s_0^{(2)}\}$ are determined by minimizing
\begin{gather}
\chi^2_\Lambda \left(s_0^{(1)}\right)=\sum\limits_{\hat s}
\left(\frac{ \Lambda_{\rm fit}\left(s_0^{(1)}\right)}
 {\lambda_s\left(\hat s,\tau,s_0^{(1)}\right)}-1\right)^2\,,\quad
\chi^2_{\Lambda'}\left(s_0^{(2)}\right)=\sum\limits_{\hat s}
\left(
\frac{ \Lambda'_{\rm fit}\left(s_0^{(2)}\right)}
{\lambda'_s\left(\hat s,\tau,s_0^{(2)}\right)}
-1
\right)^2
\,.
\label{chi_squared}
\end{gather}
We define a measure of universality $\Delta=\left\vert 
   \frac{ \Lambda_\kappa-\Lambda_a}{ \Lambda_\kappa+\Lambda_a}
    \right \vert 
    +
  \left\vert 
   \frac{ \Lambda'_\kappa-\Lambda'_a}{ \Lambda'_\kappa+\Lambda'_a}
    \right \vert$
to evaluate each resonance model; the smaller the value of $\Delta$, the better the results adhere to the principle of universality. The results of the optimization procedure are given in Table~\ref{scale_factor_table} for each channel and model, using the central values of the QCD parameters. 
\begin{table}[htb]
\centering
\renewcommand{\arraystretch}{1.1}
\begin{tabular}{|c|c|c|c|c|c|c|c|c|}
\hline
Model & Channel &   $s_0^{(1)}$ & $s_0^{(2)}$ & $\Lambda$ & $\Lambda'$  & $\chi^2_\Lambda\times 10^6$ & $\chi^2_{\Lambda'} \times 10^6$ & $\Delta$ \\
\hline
\hline
\multirow{2}{*}{PROP} 
& $K_0^*$ & $2.76$ & $1.79$ & $0.1256$ & $0.2798$ 
& $17.9$
&  $19.6$ 
& \multirow{2}{*}{$0.0586$}
\\
\hhline{|~|-|-|-|-|-|-|-|~|}
  & $a_0$  & $2.40$ & $2.13$ & $0.1169$ & $0.2928$  
 & $27.1$
 &  $6.83$  
 &
 \\
 \hline\hline
 \multirow{2}{*}{DBW} 
& $K_0^*$ & $2.90$ & $2.11$ & $0.1284$ & $0.2973$ 
& $14.4$
&  $21.0$ 
& \multirow{2}{*}{$0.0397$}
\\
\hhline{|~|-|-|-|-|-|-|-|~|}
 & $a_0$  & $2.49$ & $2.20$ & $0.1191$ & $0.2960$  
 & $27.0$ 
 &  $7.33$ 
 &
 \\
 \hline\hline
  \multirow{2}{*}{GBW} 
& $K_0^*$ & $2.97$ & $2.30$ & $0.1297$ & $0.3070$ 
& $15.6$ 
&  $28.5$ 
& \multirow{2}{*}{$0.0539$}
\\
\hhline{|~|-|-|-|-|-|-|-|~|}
& $a_0$  & $2.53$ & $2.24$ & $0.1201$ & $0.2978$  
 & $29.1$ 
 &  $8.26$  
 &
 \\
 \hline\hline
  \multirow{2}{*}{EDBW} 
& $K_0^*$ & $3.20$ & $2.34$ & $0.1339$ & $0.3091$ 
& $6.10$
&  $14.2$ 
& \multirow{2}{*}{$0.0182$}
\\
\hhline{|~|-|-|-|-|-|-|-|~|}
& $a_0$  & $3.08$ & $2.65$ & $0.1318$ & $0.3156$  
 & $13.3$ 
 &  $6.56$ 
 &
 \\
 \hline\hline
  \multirow{2}{*}{EGBW} 
& $K_0^*$ & $3.31$ & $2.49$ & $0.1358$ & $0.3163$ 
& $5.28$ 
&  $18.3$ 
& \multirow{2}{*}{$0.0119$}
\\
\hhline{|~|-|-|-|-|-|-|-|~|}
& $a_0$  & $3.14$ & $2.69$ & $0.1330$ & $0.3173$  
 & $13.2$ 
 &  $7.04$  
 &
 \\
 \hline
\end{tabular}
\caption{
Optimized and fitted values for the continuum thresholds and scale factors  are given for each model and channel, along with the (dimensionless) values  of $\{\chi_\Lambda^2,\chi_{\Lambda'}^2\}$ and $\Delta$ that quantify the scale factor properties of energy-independence and measure of universality.  All units in GeV, except for the dimensionless quantities $\chi^2$ and $\Delta$.
Resonance models are ordered in increasing sophistication and increasing width effect.}
\label{scale_factor_table}
\end{table}
Comparing to the previous benchmark analysis \cite{CLQCDSR_2020}, the PROP model under our revised analysis shows an effect-size of $\approx10\,\mathrm{MeV}$, with scale factor values increasing slightly in the revised analysis. This quantifies the difference between the two methodologies, demonstrating that it is similar to the variations seen between different resonance models in Table \ref{scale_factor_table}. We can see improvement in the measures of scale-factor energy-independence $\chi^2$ and our measure of universality $\Delta$ as the model sophistication and the model width effects increase. We also see  better separation between the two continuum values $\{s_0^{(1)},s_0^{(2)}\}$ and the resonance mass scales, suggesting a refinement of our benchmark PROP model.

\begin{figure}[hbt]
\centering
\includegraphics[scale=0.5]{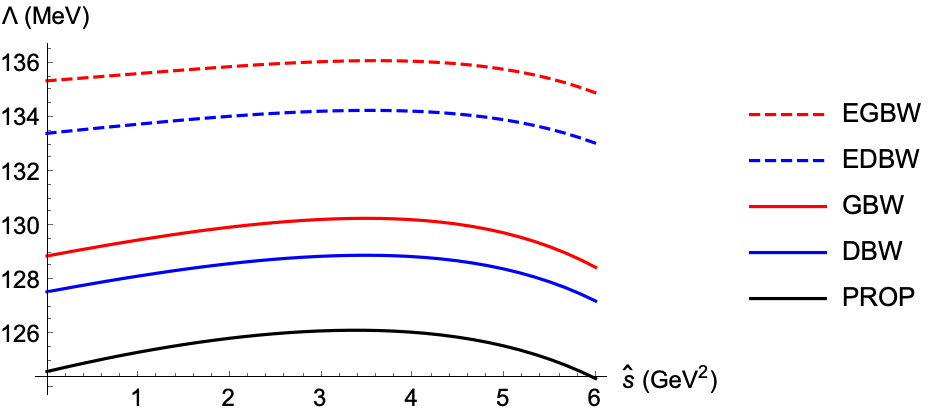}
\includegraphics[scale=0.5]{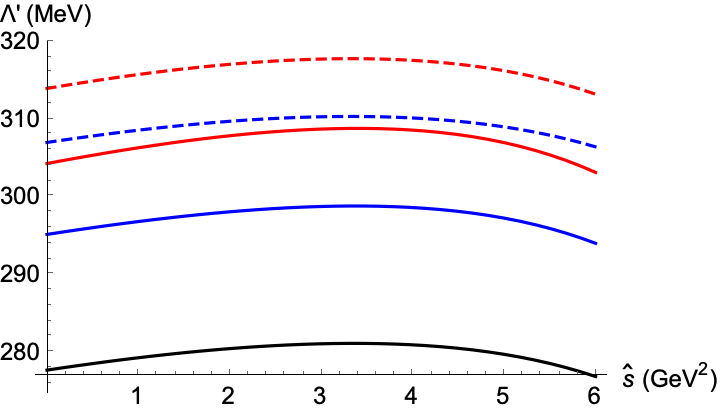}
    \caption{Theoretical predictions $\{\lambda_\kappa, \lambda'_\kappa\}$  for the scale factors [see Eqs.~\eqref{scale_relation_lambda} and \eqref{scale_relation_lambda_prime}] are shown as a function of $\hat s$
    in the isodoublet channel for the models and continuum values in Table~\ref{scale_factor_table}. The benchmark analysis value $\tau=3\,{\rm GeV^4}$ and central values of the QCD parameters have been used. 
    Scale has been chosen to highlight the differences between the models.
    }     
    \label{isodoublet_scale_factor_models_fig}
\end{figure}

\begin{figure}[hbt]
\centering
\includegraphics[scale=0.5]{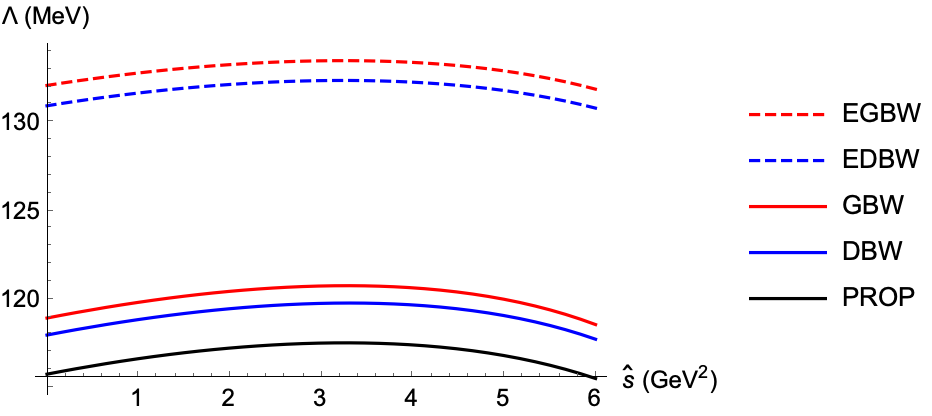}
\includegraphics[scale=0.5]{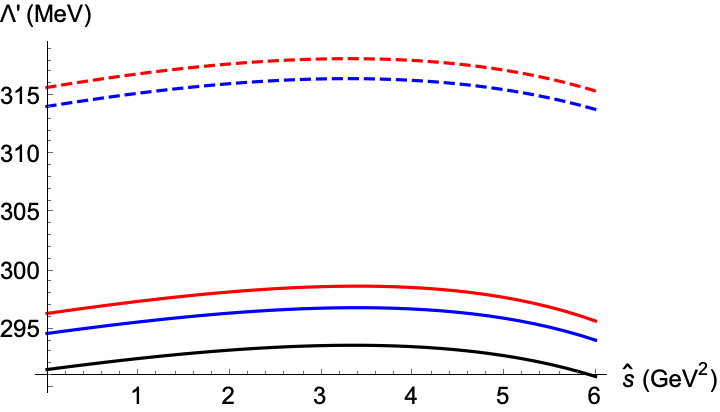}
    \caption{Theoretical predictions $\{\lambda_a, \lambda'_a\}$  for the scale factors [see Eqs.~\eqref{scale_relation_lambda} and \eqref{scale_relation_lambda_prime}] are shown as a function of $\hat s$
    in the isotriplet channel for the models and continuum values in Table~\ref{scale_factor_table}. The benchmark analysis value $\tau=3\,{\rm GeV^4}$ and central values of the QCD parameters have been used. 
    Scale has been chosen to highlight the differences between the models.
    }  \label{isotriplet_scale_factor_models_fig}
\end{figure}

Comparing the resonance models in Figures \ref{isodoublet_scale_factor_models_fig} and \ref{isotriplet_scale_factor_models_fig} for the kappa and $a_0$ systems respectively, we see a consistently increasing scale factor value as model sophistication increases, as well as an increased independence on the parameter $\hat{s}$, which is reflected in the improved values of $\chi^2$ for the extended models. Most importantly, we see that there is a clear improvement in the measure of universality with the extended models, showing the capabilities of the GSR analysis to distinguish between our different hadronic resonance models. For additional figures demonstrating the effects of the different models on the scale factors, see \cite{ExtBridge}.

We have revised our methodology connecting the chiral Lagrangian and Gaussian sum-rule methodology, providing a framework that can be extended to other isospin sectors and accommodate more sophisticated mixing effects such as those with glueballs. As well, we have developed a background-resonance interference approximation which provides an excellent descriptions of scattering cross-sections for $\pi K$ and $\pi \eta$ processes, allowing us to develop new resonance models inspired by the same approximation. These new resonance models improve our previous scale-factor analysis, leading to scale factors that better reflect the properties of energy independence and universality, and equip us with a framework for future studies of the challenging dynamics of the scalar isoscalar sector. 

\section*{Acknowledgments}
TGS acknowledges research funding from the Natural Sciences and Engineering Research Council of Canada (NSERC) and is grateful to SUNY Polytechnic Institute for hosting a sabbatical visit that initiated this work.
 %%%%%%%%%%%%%%%%%%%%%%%%%%%%%


\begin{thebibliography}{99}

\bibitem{PDG}	S. Navas et al. (Particle Data Group),  Phys. Rev. D {\bf 110}, 030001 (2024). 

\bibitem{PKZ_review}
J.R. Pelaez, Phys. Rept. {\bf 658}, 1 (2016); E. Klempt and  A. Zaitsev, 
Phys. Rept. {\bf 454}, 1 (2007).

\bibitem{GLSM}
A.H. Fariborz,  R.  Jora and J. Schechter,  Phys. Rev. D {\bf 72}, 034001 (2005);
A.H. Fariborz,  R. Jora and J. Schechter,  Phys. Rev. D {\bf 79},  074014 (2009).

\bibitem{GLSM_piK}
A.H. Fariborz,  E. Pourjafarabadi, S. Zarepour and S.M. Zebarjad, Phys.Rev. D 
{\bf 92},  113002 (2015).

\bibitem{00_BFS61}
D. Black, A.H. Fariborz and J. Schechter, Phys. Rev. D {\bf 61}, 074030 (2000).

\bibitem{GLSM_pieta}
A.H. Fariborz, S. Zarepour, E. Pourjafarabadi and S.M. Zebarjad,
%``Chiral nonet mixing in $\pi \eta $ scattering,''
Eur. Phys. J. C {\bf 82}, 1133 (2022).

  \bibitem{Parganlija:2012fy} 
  D.~Parganlija, P.~Kovacs, G.~Wolf, F.~Giacosa and D.~H.~Rischke,
  %``Meson vacuum phenomenology in a three-flavor linear sigma model with (axial-)vector mesons,''
  Phys.\ Rev.\ D {\bf 87}, 014011 (2013).

\bibitem{GLSM_pipi}
A.H. Fariborz, R. Jora, J. Schechter and M.N. Shahid,
%``Chiral Nonet Mixing in pi pi Scattering,''
Phys. Rev. D {\bf 84}, 113004 (2011).

\bibitem{GLSM_inst}
A.H. Fariborz, R. Jora and J. Schechter, Phys. Rev. D {\bf 77}, 094004 (2008). 

\bibitem{Giacosa:2006tf} 
 F.~Giacosa,
  %``Mixing of scalar tetraquark and quarkonia states in a chiral approach,''
  Phys.\ Rev.\ D {\bf 75}, 054007 (2007).

\bibitem{00_BFMNS}
D. Black,  A.H. Fariborz, S. Moussa, S. Nasri and  J. Schechter, Phys.Rev.D {\bf 64}, 014031 (2001). 

\bibitem{NLCL_kappa}
D. Black, A. H. Fariborz, F. Sannino and J. Schechter,  Phys. Rev. D {\bf 58}, 054012 (1998); D. Black, A. H. Fariborz, F. Sannino and J. Schechter, Phys. Rev. D {\bf 59}, 074026
(1999).

\bibitem{Carter:1995zi} 
  G.~W.~Carter, P.~J.~Ellis and S.~Rudaz,
  %``An Effective Lagrangian with broken scale and chiral symmetry: 2. Pion phenomenology,''
  Nucl.\ Phys.\ A {\bf 603}, 367 (1996)
  Erratum: [Nucl.\ Phys.\ A {\bf 608}, 514 (1996)].

 \bibitem{Ko:1994en} 
  P.~Ko and S.~Rudaz,
  %``Phenomenology of scalar and vector mesons in the linear sigma model,''
  Phys.\ Rev.\ D {\bf 50}, 6877 (1994).

  \bibitem{SVZ}
M.A.\ Shifman, A.I.\ Vainshtein and V.I.\ Zakharov, Nucl.\ Phys.\ B {\bf 147}, 385 (1979);
M.A.\ Shifman, A.I.\ Vainshtein and V.I.\ Zakharov, Nucl.\ Phys.\ B {\bf 147}, 448 (1979).

\bibitem{Reinders:1984sr}
  L.~J.~Reinders, H.~Rubinstein and S.~Yazaki,
  %``Hadron Properties from QCD Sum Rules,''
  Phys.\ Rept.\   {\bf 127}, 1 (1985).

  \bibitem{Narison:2002woh}
S.~Narison,
``QCD as a Theory of Hadrons: From Partons to Confinement'',
Camb. Monogr. Part. Phys. Nucl. Phys. Cosmol. \textbf{17}, 1--812 (2007)
Oxford University Press, 2007.

\bibitem{CLQCDSR_2016}
  A.~H.~Fariborz, A.~Pokraka and T.~G.~Steele,
  %``Connections between chiral Lagrangians and QCD sum-rules,''
  Mod.\ Phys.\ Lett.\ A {\bf 31}, 1650023 (2016).


\bibitem{CLQCDSR:2019_Proc}
A.~H.~Fariborz, J.~Ho, A.~Pokraka and T.~G.~Steele,
%``The Bridge Between Chiral Lagrangians and QCD Sum-Rules,''
Nucl. Part. Phys. Proc. \textbf{309-311}, 119-123 (2020)

\bibitem{CLQCDSR_2020}
A.~H.~Fariborz, J.~Ho and T.~G.~Steele,
%``Universal scale factors relating mesonic fields and quark operators,''
Mod. Phys. Lett. A {\bf 35}, 2050173 (2020).

\bibitem{ExtBridge}
  A.~H.~Fariborz, J.~Ho and T.~G.~Steele,
  %Extending the Bridge Connecting Chiral Lagrangians and QCD Gaussian Sum-Rules for Low-Energy Hadronic Physics
  Phys.\ Rev.\ D {\bf111}, 094023 (2025)

\bibitem{Chen:2007xr}
H.~X.~Chen, A.~Hosaka and S.~L.~Zhu,
%``QCD sum rule study of the masses of light tetraquark scalar mesons,''
Phys. Lett. B \textbf{650}, 369-372 (2007); H.~X.~Chen, A.~Hosaka and S.~L.~Zhu,
%``Light Scalar Tetraquark Mesons in the QCD Sum Rule,''
Phys. Rev. D \textbf{76}, 094025 (2007)

\bibitem{gauss} R.A.\ Bertlmann, G.\ Launer, E.\ de Rafael, Nucl.\ Phys.\ B {\bf 250}, 61 (1985).

\bibitem{harnett_quark} G.\ Orlandini, T.G.\ Steele, D.\ Harnett,
Nucl.~Phys.~A {\bf 686},  261 (2001).

\bibitem{Aston}D. Aston, et al., Nucl. Phys. B 296, 493 (1988).

\bibitem{Zhang:2009qb}
  J.~Zhang, H.~Y.~Jin, Z.~F.~Zhang, T.~G.~Steele and D.~H.~Lu,
  %``Light scalar mesons in QCD sum rules with inclusion of instantons,''
  Phys.\ Rev.\ D {\bf 79}, 114033 (2009). 

\bibitem{Du:2004ki}
  D.~S.~Du, J.~W.~Li and M.~Z.~Yang,
  %``Mass and decay constant of I = 1/2 scalar meson in QCD sum rule,''
  Phys.\ Lett.\ B {\bf 619},  105 (2005).

\bibitem{Jamin:1992se}
M.~Jamin and M.~Munz,
%``Current correlators to all orders in the quark masses,''
Z. Phys. C \textbf{60}, 569-578 (1993)

\bibitem{Narison:2011rn}
  S.~Narison,
  %``Gluon Condensates and m_b(m_b) from QCD-Exponential Moments at Higher Orders,''
  Phys.\ Lett.\ B {\bf 707},  259 (2012).
  
\bibitem{Belyaev:1982sa}
  V.~M.~Belyaev and B.~L.~Ioffe,
  Sov.\ Phys.\ JETP {\bf 56}, 493 (1982).

\bibitem{Beneke:1992ba}
  M.~Beneke and H.~G.~Dosch,
  %``Flavor dependence of the mixed quark gluon condensate,''
  Phys.\ Lett.\ B {\bf 284}, 116 (1992).

\end{thebibliography}
\end{document}